\documentclass[12pt, oneside]{article}

\usepackage{subfiles}           
\usepackage{standalone}         
\usepackage[utf8]{inputenc}     
\usepackage[T1]{fontenc}        
\usepackage[english]{babel}     
\usepackage{xcolor}             
\usepackage{amsmath}            
\usepackage{amssymb}            
\usepackage{gensymb}            
\usepackage[english]{babel}     
\usepackage{amsthm}             
\usepackage{hyperref}           
\usepackage{float}              
\usepackage{wrapfig}            
\usepackage{fancyhdr}           
\usepackage{geometry}           
\usepackage{changepage}         
\usepackage{mathtools}          
\usepackage{tikz, tikz-3dplot}  
\usepackage{pgfplots}           
    \usepgfplotslibrary{polar}  
    \pgfplotsset{compat=1.16}       
\usepackage{subcaption}         
\usepackage{graphicx}           
\usepackage{multicol}           
\usepackage{multirow}           
\usepackage{xfrac}              
\usepackage{listings,lstautogobble}     
    \usepackage[mono,extrasp=0em]{inconsolata}  
\usepackage{bm}                 
\usepackage{titlesec}           

    \usepackage[style=iso-numeric]{biblatex}
\definecolor{lstIdC}{HTML}{236394}
\definecolor{lstKeyC}{HTML}{0a0a0a}
\definecolor{lstComC}{HTML}{029727}
\definecolor{lstStrC}{HTML}{F44336}

    \newtheoremstyle{exampstyle}
        {0pt} 
        {8pt} 
        {} 
        {} 
        {\bfseries} 
        {:} 
        {1em} 
        {} 
    \newtheoremstyle{defstyle}
        {0pt} 
        {8pt} 
        {\normal} 
        {} 
        {} 
        {{\bfseries :}} 
        {1em} 
        {{\bfseries\thmname{#1} \thmnumber{#2} }\textit{(\thmnote{#3})}} 
    \newtheoremstyle{sectstyle}
        {8pt} 
        {8pt} 
        {\Large} 
        {} 
        {} 
        {{\bfseries :}} 
        {1em} 
        {} 
    \newtheoremstyle{subsectstyle}
        {8pt} 
        {8pt} 
        {\large} 
        {} 
        {} 
        {{\bfseries :}} 
        {1em} 
        {} 
  \theoremstyle{exampstyle}  
  \theoremstyle{defstyle}    
  \theoremstyle{exampstyle} 
  \theoremstyle{exampstyle} 
  \theoremstyle{defstyle}   
  \theoremstyle{exampstyle} 
  \theoremstyle{exampstyle} 
  \theoremstyle{exampstyle} 
  \theoremstyle{exampstyle} 
  \theoremstyle{exampstyle} 
      \theoremstyle{exampstyle} 
      \theoremstyle{defstyle}   
      \theoremstyle{exampstyle} 
      \theoremstyle{exampstyle} 
      \theoremstyle{defstyle}   
      \theoremstyle{exampstyle} 
      \theoremstyle{exampstyle} 
      \theoremstyle{exampstyle} 
      \theoremstyle{exampstyle} 
      \theoremstyle{exampstyle} 
    

\newcounter{parentnumber}

\titleclass{\subsubsubsection}{straight}[\subsection]

\newcounter{subsubsubsection}[subsubsection]
\renewcommand\thesubsubsubsection{\thesubsubsection.\arabic{subsubsubsection}}

\titleformat{\subsubsubsection}
  {\normalfont\normalsize\bfseries}{\thesubsubsubsection}{1em}{}
\titlespacing*{\subsubsubsection}
{0pt}{3.25ex plus 1ex minus .2ex}{1.5ex plus .2ex}

\makeatletter
\renewcommand\paragraph{\@startsection{paragraph}{5}{\z@}%
  {3.25ex \@plus1ex \@minus.2ex}%
  {-1em}%
  {\normalfont\normalsize\bfseries}}
\renewcommand\subparagraph{\@startsection{subparagraph}{6}{\parindent}%
  {3.25ex \@plus1ex \@minus .2ex}%
  {-1em}%
  {\normalfont\normalsize\bfseries}}
\def\toclevel@subsubsubsection{4}
\def\toclevel@paragraph{5}
\def\toclevel@paragraph{6}
\def\l@subsubsubsection{\@dottedtocline{4}{7em}{4em}}
\def\l@paragraph{\@dottedtocline{5}{10em}{5em}}
\def\l@subparagraph{\@dottedtocline{6}{14em}{6em}}
\makeatother

\def\checkmark{\tikz\fill[scale=0.4](0,.35) -- (.25,0) -- (1,.7) -- (.25,.15) -- cycle;}    

\makeatletter
\newcommand*\dotp{\mathpalette\bigcdot@{.5}}
\newcommand*\bigcdot@[2]{\mathbin{\vcenter{\hbox{\scalebox{#2}{$\m@th#1\bullet$}}}}}
\makeatother

\newcommand{\mat}[1]{\ensuremath{\boldsymbol{#1}}}                       

\renewcommand{\c}[2][]{\ensuremath{\text{\lstinline[#1]{#2}}}}      
\def\`#1'{\c[language=python]{#1}}

\def\biblio{\newpage\bibliographystyle{plain}\bibliography{comm/genbib}}
\def\subfileSetCounter#1#2{\setcounter{#1}{#2}}

\title{\yourProj}
\author{\yourName}
\date{\yourDate}

\makeatletter
\renewcommand{\@maketitle}{%
    \begin{flushleft}
        {\itshape \yourSchoolABV\, - \yourSemester\, - \yourCourseID} 
            \hfill 
            \textbf{\large \@author}
            \\[2pt]
        {\large \itshape \hspace{1em} \yourCourseTitle}
            \hfill 
        {\small\itshape (\yourID)}
            \par
        \vspace{8pt}
        \LARGE\itshape \yourProjID
        \\[6pt]
        \textbf{\Huge \@title}
        \\[-6pt]\noindent\rule{\textwidth}{0.4pt}
    \end{flushleft}
    \vspace{-28pt}
    \begin{flushright}
      {\normalsize \@date}
    \end{flushright}
    \par
}
\makeatother
\makeatletter
\newcommand{\makesubtitle}{%
    \vspace*{\fill}
    \begin{tabular*}{.85\textwidth}{rr}
      \textbf{Author:} & \yourName \\
      \textbf{Mentor:} & \yourProfName \\ 
      \textbf{Course Supervisor:} & \courseSuprivisorName \\
      \textbf{\yourSchoolABV~\! SoC Director of Undergraduate Research:} & \DUGname \\
    \end{tabular*}
}
\makeatother

\def\yourName{Andrew Osterhout} 
\def\yourID{u1317172} 

\def\yourSchoolABV{UofU} 
\def\yourSemester{Sp2022} 
\def\yourCourseID{CS 4970} 
\def\yourCourseTitle{Bachelor's Thesis} 
\def\yourCourseShortTitle{BS Thesis} 

\def\yourProfName{Ganesh Gopalakrishnan PhD} 
\def\courseSuprivisorName{Marina Kogan PhD} 
\def\DUGname{H. James de St. Germain PhD} 

\def\yourProj{SIMT/GPU Data Race Verification using ISCC and Intermediary Code Representations: A Case Study} 
\def\yourProjID{Batcheoler's Thesis} 

\def\yourDate{\today} 

\begin{document}
    
  \def\biblio{ }
  \def\subfileSetCounter#1#2{ }

  \pagenumbering{gobble}

  \maketitle


  \makesubtitle

  \clearpage    
  \pagenumbering{roman}


\subsection*{Abstract}
\addcontentsline{toc}{section}{Abstract}
\markright{Abstract}

It is often difficult to write code that you can ensure will be executed in the right order
when programing for parallel compute tasks.
%
Due to the way that today's parallel compute hardware,
primarily Graphical Processing Units (GPUs), 
allows you to write code that needs to be executed
on more threads than the device can run in parallel at once,
it is easy to write code that may result in one thread modifying data before another thread gets to use it
(this is part of what some in parallel programing call ``memory model rule violations''),
thus confounding the calculations.

It would be of great help to be able to have a tool that could verify that the code will execute 
as expected.
However, with many languages that have support for performing operations in a parallel compute environment,
all of these must at some point compile the code to run on the GPUs
making it very difficult to support every language and every hardware type.
Additionally most of these programing languages 
are designed for normal single threaded and concurrent programing, 
and just use libraries and other tools to transform the code into something supported by parallel compute hardware.
Which is unlikely to change when parallel compute programing is used by comparatively few programers
compared to programers using the more common single-threaded and concurrent programing paradigms.
%
This is largely due to the differences in tasks that the different paradigms specialize in,
and that most of these parallel compute tasks need to be overseen by control code that use these other paradigms.
So it makes sense, 
for convenience and organization, 
to allow you to easily code such algorithms in the same language and files as the rest of your code.

Therefore, it would be of great use to be able to perform verification and analysis on the Memory Model 
of a parallel compute code, 
in the lower level intermediary representations that the code goes through during compilation.
Because, 
many of the different starting languages and ending representations in the compilation process 
go through the same intermediary representations along the way:
meaning that if you could perform it at a LLVM-IR level (for languages that use the LLVM compilation path),
the PTX Intermediary code (for CUDA compatible GPUs), 
or possibly even the assembly language the hardware uses (for NVIDIA GPUs it is called SASS),
you could then support far more languages, hardware configurations, and compilation pathways,
the further into into the compilation process where many compilation paths must converge,
one operates at.

%
This body of work aims to deal with the question of if 
there is
still enough of the information in the intermediary representations 
to be able to perform memory model verification to check for data races.
Such a verification analysis would be part of the larger work that this work is a part of.  
%
%
This is a crucial thing to know as during compilation optimizations 
and changing the grammar used to represent the actions/meaning 
loose the human understandable shape that most programming methods use.
%
For this we plan to analyze as a case study the Ge-SpMM Sparse Matrix Multiplication Algorithm \cite{huang2020},
implemented in CUDA C++ with the LLVM compiler and Julia with CUDA.jl.

\begin{table}[b]
  \centering
  \begin{tabular}{|r|c|}
    \hline
    \multicolumn{2}{|c|}{Project Resources} \\\hline
    \textbf{Thesis Presentation Slides:} & \href{https://bit.ly/oster-bs-thesis-presentation}{bit.ly/oster-bs-thesis-presentation} \\
    \textbf{Scratch Code Repo:} & \href{https://github.com/osterhoutan-UofU/Ge-SpMM_Julia_test}{github.com/osterhoutan-UofU/Ge-SpMM\textunderscore~Julia\textunderscore~test} \\
    \textbf{Ge-SpMM Analysis Presentation:} & \href{https://bit.ly/oster-Ge-SpMM-presentation}{bit.ly/oster-Ge-SpMM-presentation} \\ \hline
  \end{tabular}
  \caption{Links to Project Resource}
  \label{tbl:proj-resources}
\end{table}


  
  \newpage
  \markright{Contents}
  \addcontentsline{toc}{section}{Contents}
  \tableofcontents

  \clearpage
  \pagenumbering{arabic}
  

\section{Introduction}


In Parallel Programing it is important to write code,
that despite running multiple threads using a shared pool of memory, 
will read and write to said memory in an order that would not break the 
``big picture'' procedural contract with a verified linear/single-threaded implementation of the algorithm.
A concept called ``Data Dependency,''
since it it describes how some parts of the data are dependent on the completion of other parts of the data,
%
%
but more importantly that sometimes you might have one thread that will finish it's work and 
write the result to shared memory, on top of a piece of data that another thread still needed to complete its task,
a situation commonly referred to as a ``Data Race''.

The best solution to this is to write your code such that this doesn't happen.
Which can be difficult,
but there are solutions out there that can help.
Like Michael Kruse et al.'s Polyhedral Parallel Code Generation (PPCG) tool \cite{kruse2021}.
Which can be used to take normal linear/single-threaded algorithms written in C or limited C++
\footnote{The use of all but a few simple standard libraries is not supported on most GPUs, 
           so make sure not to use them.}
and produce the kernel code, in either CUDA or OpenCL,
including intelligent launch sizing for threads and shared memory,
and proper thread fencing in the kernel code.
%

There are also tools that allow you to analyze your algorithm in a more manual but mathematical way.
Through the use of Integer Sets
\footnote{This is the same underlying technique used for the PPCG tool, 
          but with a different approach for how humans interact with the technology} 
and an ``interactive calculator'' environment, 
provided \textit{via} integrations with various programing languages like python 
and a custom DSL
\footnote{Domain Specific Language}
\cite{verdoolaege2013,verdoolaege,verdoolaege2014,verdoolaege2016,verdoolaege2016a}.
It is possible to analyze how data is transferred from one memory location to another,
when you consider every variable an array that can be represented as a set of its index values.

This integer set approach seems fundamental to how memory model verification in terms of data dependencies can be approached.
It is our hope that it can be adapted to work to analyze the lower level representations 
that the intermediary representations use,
by representing the various registers, ``pointers'' and temporary variables as sets of their own.
However, our success will depend on if in these semi-compiled states the, 
change from programers original code representation,
to the more computer compatible 
intermediary code formats, such as LLVM-IR, and PTX still contain enough identifiable information to 
even consider extending the integer set analysis to operate on their ever shifting ``pointer'' representation.

In addition to the basic data handling, we will also need to be able to identify 
some static components in the intermediary representations,
so that we can ensure that we accurately replicate the algorithm for our future verification system.
(1) calls to the driver and hardware,
to do things like retrieve the thread id, used to figure out what part of the work a thread will do;
(2) where the calls to thread fences and other barriers are;
(3) how shared memory allocation, references and reads and writes are handled;
and (4) where the input and output data locations will end up.



%
To accomplish these goals
we intend to perform a case study on Huang et al's 
General-purpose Sparse Matrix Multiplication (GeSpMM) algorithm \cite{huang2020}.
We chose this algorithm because, we know it is a good algorithm that by its nature of treating the input matrices as immutable
and therefore will not have any data dependencies to throw off our analysis,
but does utilize a great deal of the strategies for parallel code optimization that have to be done 
by the programer level.
Such as, the use of retrieving data from global memory and storing it in the faster shared memory,
by means of split work memory retrieval between threads.

It must also be noted that we will only be working with technologies that can be used with NVIDIA's CUDA ecosystem. 
To that effect we intend to examine the LLVM IR, CUDA PTX (ISA) and possibly even CUDA SASS, 
and compatible pipelines.
Primarily the LLVM CUDA compilation pathway, that can take any language that has a pathway to LLVM IR or can be compiled with an LLVM tool, and compile the LLVM IR to PTX and/or SASS 
by a hybrid mixture of direct translation and the use of tools available in the CUDA toolkit.
%
%
To this effect, we will utilize the pathway from C/C++ with LLVM's clang compiler to get or intermediary representations if LLVM IR and PTX.
Then we will use the CUDA.jl package\cite{juliagpugroup2021} for Julia Lang\cite{bezanson2012} (a language built with LLVM technologies) to a
allow us to compile our kernel codes to LLVM IR and PTX to use in our case study.
This decision means that we will not be looking into the other primary parallel compute ecosystem surrounding the OpenCL toolkit made by the Khronos Group\cite{khronosgroup2013}, 
or any other parallel compute ecosystem.



\subsection{Background}


There is a good amount of work in the fields of parallel algorithm data model verification,
as well as in various fields concerning integer set representations of memory models in algorithms.
Both of which are important for this body of work 
in addition to the larger verification process that we intend to do.
We will discuss some of these concepts and supporting work in this section,
followed by some supporting work for tools that we looked into to use
and work pertaining to underlying concepts such as de-compilation of NVIDIA CUDA SASS.
Finally, we will go into further detail about the Ge-SpMM \cite{huang2020} algorithm 
we are using as the subject of our case study.

\subsubsection{Foundational Work}

The primary efforts made in the field of parallel algorithm code generation and analysis 
that are still current enough to be considered more than inspirational,
are all implementations of the Integer Set Library.
Namely the PPCG\cite{kruse2021} and iscc\cite{verdoolaege2013} projects,
where through the use of integer sets, representing arrays, and transferring elements around,
they are able to produce orderings of operations that do not cause data dependencies.

In Verdoolaege et al.'s 2014 presentation/tutorial on their iscc tool\cite{verdoolaege2014},
they explain how they utilize integer sets using the 
Integer Set Library (ISL)\cite{kruse2021a},
to find ``temporal dependencies'' 
that need to be preserved when performing their loop-tiling and unrolling.
%
%
More important to our research they indirectly show us what a 
parallel memory model can like
when represented with integer sets representing the ``high level'' code representations
(primarily C) representations look.
At least what they look like when they are the result of their polyhedral code transformations,
but this should be more than sufficient for our purposes,
since we just need to see how arrays are represented and how they differentiate the varies possible paths 
of the parallel thread operations.


\subsubsection{Related Technologies}

There is also much work and documenting for both of the main intermediary representations 
we intend to work with 
(LLVM IR \cite{llvmfoundation2021} and NVida CUDA PTX (ISA) \cite{nvidiacoorperation2021}).
Though it should be noted that LLVM IR is not specifically meant for parallel execution code,
but is a common step in compilation to get kernel codes.

Both LLVM IR and PTX, utilize a syntax that breaks down any complex statements into single line instructions,
that can be more easily mapped to assembly/binary instruction representations.
They both utilize a system of ``identifiers'' that work similarly to variables, 
but like machine level registers can only store integers, pointers and floating point values.
Of even greater importance they both allow some way to represent ``global'' memory,
functions declarations with basic typing info, registers (not just variable identifiers)
and support for modules in identifier names.
This form of intermediary representation allows you to preserve a good amount of the high level details, including many aspects of naming schemes and module structure, 
but gets rid of any order of operations and clever tricks of code writing that allow for more readable code.

LLVM IR and PTX do differ in several key aspects: 
(1) LLVM IR, has some limited support for better representations of object oriented programing code,
such as inheritance, class method membership and data value struct representations,
but PTX has only support for the struct representation of a type;
(2) LLVM IR uses a more friendly syntax of: ``\lstinline{
while PTX uses a more traditional assembly syntax form: ``\lstinline{<instruction> <param> ... ;}'';
and (3) many more not mentioned here.
%
%

While I only plan to look at NVIDIA's CUDA ecosystem
for reasons pertaining to the scope of the larger project this thesis is a part of,
it might be possible to adapt any findings found with LLVM IR to the OpenCL IR 
of the OpenCL project, which is derived from LLVM IR\cite{khronosgroup2013}. 

\subsubsection{Useful Tools}


To help supplement my efforts there are also some great resources for understanding and working with
these low level GPU code,
In the form of a new tool called NVBit\cite{villa2019}, that can read and inject code into kernels at run time,
allowing you to see without making too many changes to the low level SASS what is happening.
As well as some work done in recompiling/decoding CUDA code, 
from Hayes et al.\cite{hayes2019} that may prove useful as well.



\subsubsection{About Ge-SpMM}

The algorithm we will be using as our case study is the General-purpose Sparse Matrix Multiplication (Ge-SpMM) algorithm from Huang et al.'s 2020 paper published in arXiv 2020
conference journal entitled:
\textit{``GE-SpMM: General-purpose Sparse Matrix-Matrix Multiplication on GPUs for Graph Neural Networks''}
\cite{huang2020}.
Which as the title implies is a ``general-purpose'' sparse matrix multiplication algorithm,
that is optimized for graph neural networks.
It utilizes the Compressed Sparse Row (CSR)\cite{eisenstat1977} format 
for compressing the \mat{A} matrix in the matrix multiplication form:
$\mat{A}\times\mat{B}=\mat{C}$.
Where \mat{A} is used as an ``in-edge adjacency matrix'' used on an ``input feature matrix'' \mat{B}
to produce ``output feature matrix'' \mat{C}.

Their results for their intended goals seem promising, but we are more interested in their algorithm
because of the features they implement to optimize the algorithm, for use with GPUs.
%
%
These features of interest to us include:
(1) \textit{``Coalesced Row Caching (CRC),''} 
where shared memory is used to allow parallel threads working on the same row of the sparse matrix to fetch a single value into adjacent portions of shared memory that all threads will need to read at least once from the slower global memory;
and (2) \textit{``Course Grained Warp Merging (CWM),''} 
where they reduce thread/warp synchronization overhead by coding to allow the use of warp level 
thread synchronization rather than block level reducing the time each thread needs to wait to continue.
These features are of interest to us as they utilize the use of shared memory and thread synchronization
via driver/device level interfaces.
Which are important to be able to identify in any intermediary representation we work with for 
data race verification in the future.

Additionally, this Ge-SpMM algorithm is written in such a way that it does not contain any data races.
Therefore, we should be able to use it and perform the integer set analysis without any issues that could
arise from data races confounding the transfer and manipulation of the ``data'' in the analysis.


  

  \newpage

\section{Methods}


For this project as partially described in the previous sections,
I intend to via a case study with the Ge-SpMM\cite{huang2020} algorithm
to discover if it is possible to trace the memory model of a parallel algorithm
represented in the intermediary forms of LLVM IR and CUDA PTX (ISA) in a way that 
would be meaningful to our future goal of a memory model data dependency verification tool.
The Ge-SpMM team has graciously provided sever implementations of its algorithm in CUDA C/C++,
with various levels of loop unrolling to increase the efficiency of the algorithm,
in their \href{https://github.com/hgyhungry/ge-spmm}{source repo}
\footnote{Ge-SpMM\cite{huang2020} source repo: 
          \href{https://github.com/hgyhungry/ge-spmm}{https://github.com/hgyhungry/ge-spmm}}.
I have also made an implementation of in the Julia Lang programing language\cite{bezanson2012} 
with the CUDA.jl\cite{juliagpugroup2021} 
library that enables CUDA programing in Julia.
For C/C++ implementations we will compile to LLVM IR and PTX using the Clang compiler.
Additionally, since Julia is a Just in Time (JiT) programing language implemented with LLVM
and the CUDA.jl package \cite{juliagpugroup2021} operates such that it just uses same LLVM 
provided CUDA compilation pathway 
after scrubbing away some of the unessisary Julia components from the LLVM IR representation
to get the kernel code, we can also get the requisite LLVM IR and PTX representations
from our Julia code as well,
but with some extra Julia information in at least the LLVM IR if not also the PTX representations.

This extra information, from Julia that possibly exists in the intermediary representations,
is one of the biggest reasons we chose to look at CUDA kernels compiled from Julia to begin with.
As comparing it to any code from the much more low-level CUDA-C/C++ algorithms will help us 
differentiate extra information that higher level languages might leave in their intermediary representations durning compilations.
%

From this state I will determine if there components I need can be located or at least derived 
from the intermediary code representations.
Those components being:
(1) function/kernel header information, especially as it has to do with input and output data parameters;
(2) basic typing information for arrays, pointers, floats and the various integral types;
(3) kernel/driver calls to retrieve thread, block and tile coordinate information;
(4) kernel/driver calls to any of the built in thread fencing ``functions;''
(5) kernel/driver calls to retrieve a memory location for a section of shared memory;
(6) pointers to memory locations inside any array;
and finally (7) when/where data is being combined and moved around.
To do it, first we will need to identify calls to any driver functions and reserved registers,
make sure we can identify what call and/or information it relates too.
Then we will attempt a manual data trace of the elements of the data to better our understanding,
before using the ISCC interactive calculator tool \cite{verdoolaege2013}.

If all goes well, and it seems possible that there is a way to identify all the elements we need 
to perform our analysis, and we have time we may move on to analyzing at least one known bad algorithm in
a similar manner
to see if it affects our ability to work with the intermediary representations.
But in the end we hope to have been able to better our understanding of the mechanics of parallel algorithms
and the effects of various forms code representations has on the information that can be derived
from said representations
in terms of what we need to perform our future memory model analysis on data-dependencies and data-races,
in our planned future work.


\newpage

\subsection{Using ISCC in Formal Analysis/Verification}\label{sec:iscc-formal}

For our work we will be using the ISCC tool, to construct the integer set for our \`ISLSchedule'
then test it for Read after Write Dependencies (\`RaW'), Write after Write Dependencies (\`WaW'),
and Write after Read Dependencies (\`WaR').
\textit{(See Listing \ref{lst:iscc-polyp-full} for the full ISCC ``code'' example discussed in this section.)}

For Example to do this for a simple polyhedral multiplication of the C code 
representing a basic matrix multiply seen in Listing \ref{lst:polyp-c-code},
\begin{lstlisting}[label={lst:polyp-c-code},caption={Basic Matrix Multiply C Code},numbers=left]
void polyhedral_product(int n, int *A, int *B, int *C) {
    for(int k = 0; k < 2*n-1; k++)
S:      C[k] = 0;
    for(int i = 0; i < n; i++)
        for(int j = 0; j < n; j++)
T:          C[i+j] += A[i] * B[j];
}
\end{lstlisting}
\noindent
we would do the following:
\begin{enumerate}
  \item Decide on code segments that can be treated as ``functions'' of sorts. 
        These must not intersect with the exit boundary of any loop or conditional branch structure that the segment starts in.
        In this case we can say that the operation of setting \`C[k] = 0' 
        (line 3 in Listing \ref{lst:polyp-c-code})
        can be a code section we will label as \`S' in our ISCC,
        and that the \`C[i,j]= A[i] * B[j]'
        (line 6 in Listing \ref{lst:polyp-c-code})
        can be a code section we will label as \`T' in our ISCC.

  \item Create a Domain for our schedule in terms of these code sections.
        Where we define them as labeled parameterized integer sets, and constrain their domains using presburger notation.
        \textit{(See Listing \ref{lst:iscc-polyp-domain})}.
\begin{lstlisting}[language=python,numbers=left,linerange={10-13},firstnumber=10,%
                        label={lst:iscc-polyp-domain},%
                        caption={The Domain in ISCC notation for the polyhedral product example %
                                  %\textit{(Exerpt from Listing \ref{lst:iscc-polyp-full})}.}%
                        }]
Domain := [n] -> {
    S[k] : k <= -2 + 2n and k >= 0;
    T[i, j] : i >= 0 and i <= -1 + n and j <= -1 + n and j >= 0;
};
\end{lstlisting}
        \noindent
        We allow the domain to be described by the \`n' symbol, since we have no idea the size of the Arrays we are dealing with.
        Normally, the start and end segments of a loop are enough to identify the domain restrictions and that is what we do in this example.

  \item Define a \`Read' set dependent on the same variables as the domain, 
        and constringed as ``implications'' from the domain.
        Then factor this set with the \`Domain' to get a domain respecting set of read occurrences
        \textit{(See Listing \ref{lst:iscc-polyp-read})}.
\begin{lstlisting}[language=python,numbers=left,linerange={15-19},firstnumber=15,%
                        label={lst:iscc-polyp-read},%
                        caption={The Read occurrences in ISCC notation for the polyhedral product example %
                                  %\textit{(Exerpt from Listing \ref{lst:iscc-polyp-full})}.}%
                        }]
Read := [n] -> {
    T[i, j] -> C[i + j];
    T[i, j] -> B[j];
    T[i, j] -> A[i];
} * Domain;
\end{lstlisting}
  \item Repeat the previous steps for all occurrences of writes to the various memory locations
        \textit{(See Listing \ref{lst:iscc-polyp-read})}.
\begin{lstlisting}[language=python,numbers=left,linerange={21-24},firstnumber=21,%
                        label={lst:iscc-polyp-write},%
                        caption={The Write occurrences in ISCC notation for the polyhedral product example %
                                  %\textit{(Exerpt from Listing \ref{lst:iscc-polyp-full})}.}%
                        }]
Write := [n] -> {
    S[k] -> C[k];
    T[i, j] -> C[i + j];
} * Domain;
\end{lstlisting}
  
  \item Construct a \`Schedule' to transcribe the labeled integer sets to non-labeled sets, 
        that are then constrained by an ordering variable as the first parameter.
        Thereby encoding what order the labeled code sections are expected to execute in.
        \textit{(See Listing \ref{lst:iscc-polyp-schedule})}
\begin{lstlisting}[language=python,numbers=left,linerange={26-29},firstnumber=26,%
                        label={lst:iscc-polyp-schedule},%
                        caption={The scedule for the labled code sections in ISCC notation for the polyhedral product example.
                        }]
Schedule := [n] -> {
    T[i, j] -> [1, i, j];
    S[k] -> [0, k, 0];
};
\end{lstlisting}

  \item Perform the Dependency Test as normal for ISCC.
        \textit{(see lines 34-73 of Listing \ref{lst:iscc-polyp-full} for a full example 
                  or see the ISCC tutorial citation for more information on the process \cite{verdoolaege2014})}
\end{enumerate}

\subsubsection{Problems to deal with:} \label{sec:iscc-rep-problems}

For normal code the domain restrictions are obvious and described in the \`for'-loops' headers,
and often just plug right in with constants or possibly an added variable like the \`n' in the example given above.
Some code might have a boundary that is bounded by a piece of data inside an input array.
This causes major problems as ISCC has no direct notion of this kind of relationship.
However, by broadening the domain to be larger than any problem could be we can easily 
define some boundary identifiers then constrain the counter to these values.
This will result in our domain being a super-set of the actual domain.
Therefore if it passes the dependency checks then the code must be valid,
because it operates in a subset of the domain tested in the dependency test.
However, if it fails the dependency checks there is a likelihood that this failure is a false positive,
and human eyes should look to see if the offending portions are actually possible values.

\newpage
\subsubsection{Doing this for Ge-SpMM}\label{sec:iscc-formal-ge-spmm}

To take this basic methodology and apply it to the subject of our case study Ge-SpMM \cite{huang2020},
in particular their non-loop-unrolled, shared memory optimized Algorithm 2
(seen below in Figure \ref{fig:ge-spmm-algo2}), 
\begin{figure}[H]
  \centering
  \includegraphics[width=.55\textwidth]{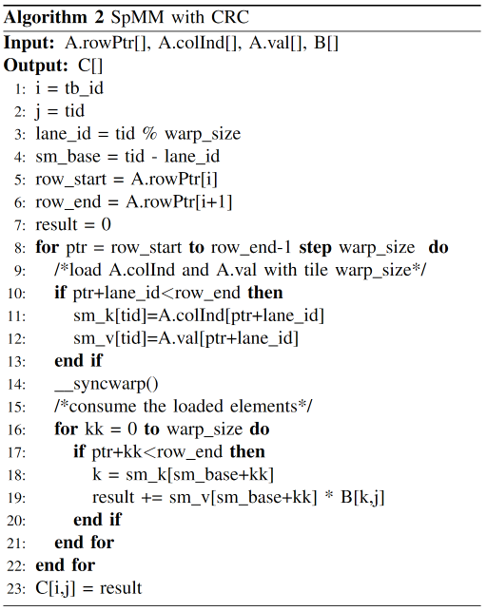}
  \caption{Ge-SpMM's Algorithm 2 \textit{\footnotesize (credit: Huang et al's 2020 paper ``GE-SpMM: \dots'' \cite{huang2020})}}
  \label{fig:ge-spmm-algo2}
\end{figure}
\noindent
we will do much the same process as before, 
but with the variations as seen bellow:
\begin{enumerate}
  \item Decide on code segments that can be treated as ``functions'' of sorts. 
        These must not intersect with the exit boundary of any loop or conditional branch structure that the segment starts in.
        In this case we can say that the operation of extracting the \`row_start'and \`row_end' values from
        the \`A.rowPtr' array
        (lines 5-6 in Figure \ref{fig:ge-spmm-algo2})
        can be a code section we will label as \`I' in our ISCC,
        and that the moving of the relevant \`A.colInd' and \`A.val' values into 
        \`sm_k' and \`sm_v' (shared memory arrays used for faster reads of the data)
        (lines 11-12 in Figure \ref{fig:ge-spmm-algo2})
        can be a code section we will label as \`F' in our ISCC,
        and finally that all of the matrix multiplication's multiplication and addition operations
        for a cell in the output matrix \`C'
        (lines 18-19 in Figure \ref{fig:ge-spmm-algo2})
        can be a code section we will label as \`T' in our ISCC representation.

  \newpage
  \item Create a Domain for our schedule in terms of these code sections.
        Where we define them as labeled parameterized integer sets, and constrain their domains using presburger notation.
        \textit{(See Listing \ref{lst:iscc-polyp-domain})}.
\begin{lstlisting}[language=python,numbers=left,linerange={5-16},firstnumber=5,%
                        label={lst:iscc-ge-spmm-domain},%
                        caption={The Domain in ISCC notation for Ge-SpMM Algorithm 2.%
                                  %\textit{(Exerpt from Listing \ref{lst:iscc-polyp-full})}.}%
                        }]
Domain := [M,N,K,A_S] -> {
    I[i, j] : 0 <= i < M and 0 <= j < N;
    F[i, j, rs, re, ptr] : 0 <= i < M 
                       and 0 <= j < N 
                       and 0 <= rs <= ptr 
                       and ptr <= re <= A_S;
    T[i, j, rs, re, ptr, kk] : 0 <= i < M 
                           and 0 <= j < N 
                           and 0 <= rs <= ptr 
                           and ptr <= re <= A_S 
                           and 0 <= kk < 32;
};
\end{lstlisting}
        \noindent
        We allow the domain to be described by the \`M', \`N', \`K' and \`A_S' symbols, 
        since we have don't the size of the Arrays/Matrices we could be dealing with.
        Normally, the start and end segments of a loop are enough to identify the domain restrictions and that is what we do in this example.
        However, this is one of the cases as mentioned in Section \ref{sec:iscc-rep-problems},
        where the loop iterator/index-variable is defined in terms of the contents of one of the input arrays/matrices,
        therefore we must add some parameters (\`rs' and \`re')
        to our integer sets to define our iterator (\`ptr') to,
        such that the domain will be a proper superset of what could actually occur in the algorithm. 

  \item Define a \`Read' set dependent on the same variables as the domain, 
        and constringed as ``implications'' from the domain.
        Then factor this set with the \`Domain' to get a domain respecting set of read occurrences
        \textit{(See Listing \ref{lst:iscc-ge-spmm-read})}.
\begin{lstlisting}[language=python,numbers=left,linerange={20-30},firstnumber=20,%
                        label={lst:iscc-ge-spmm-read},%
                        caption={The Read occurrences in ISCC notation for Ge-SpMM Algorithm 2.%
                                  %\textit{(Exerpt from Listing \ref{lst:iscc-polyp-full})}.}%
                        }]
Read := [M,N,K,A_S] -> {
    I[i, j] -> A_rowPtr[i];
    I[i, j] -> A_rowPtr[i+1];
    F[i, j, rs, re, ptr] -> A_colInd[ptr + (j % 32)] : (ptr + (j % 32)) < re;  
    F[i, j, rs, re, ptr] -> A_val[ptr + (j % 32)] :  (ptr + (j % 32)) < re;     
    T[i, j, rs, re, ptr, kk] -> sm_k[(j - (j % 32)) + kk] :  (ptr+kk) < re;
    T[i, j, rs, re, ptr, kk] -> sm_val[(j - (j % 32)) + kk] :  (ptr+kk) < re;
    T[i, j, rs, re, ptr, kk] -> B[k,j] : 0 <= k < K 
                                     and (ptr+kk) < re;
    T[i, j, rs, re, ptr, kk] -> C[i,j] : (ptr+kk) < re;
  } * Domain;
\end{lstlisting}
  
  \newpage
  \item Repeat the previous steps for all occurrences of writes to the various memory locations
        \textit{(See Listing \ref{lst:iscc-ge-spmm-read})}.
\begin{lstlisting}[language=python,numbers=left,linerange={32-36},firstnumber=32,%
                        label={lst:iscc-ge-spmm-write},%
                        caption={The Write occurrences in ISCC notation for Ge-SpMM Algorithm 2.%
                                  %\textit{(Exerpt from Listing \ref{lst:iscc-polyp-full})}.}%
                        }]
Write := [M,N,K,A_S] -> {
    F[i, j, rs, re, ptr] -> sm_k[j] : (ptr + (j % 32)) < re;
    F[i, j, rs, re, ptr] -> sm_val[j] : (ptr + (j % 32)) < re;
    T[i, j, rs, re, ptr, kk] -> C[i,j] : (ptr+kk) < re;
} * Domain;
\end{lstlisting}
  
  \item Construct a \`Schedule' to transcribe the labeled integer sets to non-labeled sets, 
        that are then constrained by an ordering variable as the first parameter.
        Thereby encoding what order the labeled code sections are expected to execute in.
        \textit{(See Listing \ref{lst:iscc-ge-spmm-schedule})}
\begin{lstlisting}[language=python,numbers=left,linerange={38-42},firstnumber=38,%
                        label={lst:iscc-ge-spmm-schedule},%
                        caption={The scedule for the labled code sections in ISCC notation for Ge-SpMM Algorithm 2.
                        }]
Schedule := [M,N,K,A_S] -> {
    I[i, j] -> [0, i, j];
    F[i, j, rs, re, ptr] -> [1, i, j, rs, re, ptr]; 
    T[i, j, rs, re, ptr, kk] -> [2, i, j, rs, re, ptr, kk];
};
\end{lstlisting}

  \item Perform the Dependency Test as normal for ISCC.
        \textit{(see lines 34-73 of Listing \ref{lst:iscc-ge-spmm-full} for a full example 
                  or see the ISCC tutorial citation for more information on the process \cite{verdoolaege2014})}
\end{enumerate}


\newpage

\subsection{Analyzing the Intermediaries}

In order to perform the ISCC data-race/dependency check, on the intermediaries,
we must first be able to extract the information needed for the steps outlined in
section(s) \ref{sec:iscc-formal} (and \ref{sec:iscc-formal-ge-spmm}).

This problem contains a few subproblems in the well researched field of decompilation,
so much of our work is based around the adaptation of these methods to suit our needs.
Luckily, we do not require full-decompilation of any of the intermediaries,
but rather just some information that decompilation extracts and uses along the way.
Such as identifying loop and conditional structures in instructional code.
Even better though is that these intermediaries still contain notions of types,
and ``modules'' for the identification of external code that has not been inlined.
We also needed a way to track where data was going as it was moved between registers,
and other memory locations.
So that we could identify when index values were being used and on what dependent data structure they were being used with
(for identifying the read and write occurrences).
To do this we employed some basic memory/expression propagation algorithms we got from 
backerstreet.com \cite{backerstreet2015}.

\noindent
The high-level algorithm that we used for extracting goes as follows:
\begin{enumerate}
  \item Identify Kernel primary function, inputs, outputs and shared data stores.
  \item Identify the outermost iterators (those derived from kernel calls to block and thread ids/sizes).
  \item Identify the Loops (their iterators and bounds).
  \item Identify the conditional branches, and kernel level syncs and fences.
  \item Using these bounds and the expression/memory propagation determine the ``code sections'' of interest to model in ISCC.
\end{enumerate}

This methodology is for the most part straight forward when using the methods from decompilation,
with the exception of the last step.
Which is far more difficult, as it requires you to exclude overhead instructions used for language overhead
(primarily a problem in LLVM).
Thankfully the memory/expression propagation information lets you figure out what is actually useful,
by checking for occurrences where memory locations of interest propagate into a dirent memory location of interest.
Then even if we factor in a few unnecessary instructions, there isn't likely to be any issues that would yield a false 
negative in the data-race/dependency check, presuming the standard compilation promises are true.





  \newpage

\section{Results of the Ge-SpMM Case Study}

Using my rough algorithms for extracting the information in the intermediaries and then creating an ISCC
representation for the intermediary code,
we were able to produce a few successful representations that we then ran as part of the standard ISCC
data-race/dependency check.

The Starter code we used was the provided Ge-SpMM CUDA C/C++ code\footnote{\url{https://github.com/hgyhungry/ge-spmm/blob/f62f51169eb26c0d4411f6d9744eb585854410e1/spmm_test.cu}}
(the \`spmm_test1' kernel function to be precise) \cite{huang2020},
compiled with clang to output plaintext LLVM IR and PTX.
As well as, a custom made Julia \cite{bezanson2012} representation of the same algorithm\footnote{\url{https://github.com/osterhoutan-UofU/Ge-SpMM_Julia_test}}
using the CUDA.jl \cite{juliagpugroup2021}, to compile and output the code as LLVM IR and PTX.

We then by hand following some basic Pseudo-code algorithms,
to prevent as much human intuition from corrupting the results as possible,
performed the information extraction and
generation of an ISCC representation.

\vspace{1em}
\noindent
The results were as follows:
\begin{table}[H]
  \centering
  \begin{tabular}{r|cc}
    Intermediary & Successful ISCC rep? & Passed Data-Race Check? \\ \hline
    \textit{Ge-SpMM Algo 2 Pseudo-code} & \checkmark & \checkmark \\
    Ge-SpMM Algo 2 CUDA C/C++$\rightarrow$\textbf{LLVM} & \checkmark & \checkmark \\
    Ge-SpMM Algo 2 CUDA C/C++$\rightarrow$LLVM$\rightarrow$\textbf{PTX} & \checkmark & \checkmark \\
    Ge-SpMM Algo 2 Julia$\rightarrow$\textbf{LLVM} & \textbf{X} & ? \\
    Ge-SpMM Algo 2 Julia$\rightarrow$LLVM$\rightarrow$\textbf{PTX} & \checkmark & \checkmark
  \end{tabular}
  \label{tbl:results}
  \caption{The results of the case study in terms of successful hand generation of ISCC representations from intermediary representations, and if the ISCC race checker provided the expected results from the data.}
\end{table}

All were successful, except the process of producing the ISCC representation the LLVM IR when compiled form Julia.
This was largely becurse the extra bloat from Julia being a JIT-ed language put to much bloat into the LLVM IR,
that would have been impractical to sift through by hand, given the time alloted for this project.
So it remains an unknown.

Because, of the nature of the Ge-SpMM algorithm outputting the results as a new matrix, and 
having one thread work on only one cell of the output matrix,
the base algorithm should be data-race free.
The only questionable part being the parallel retrieval and storage of the contents of the sparse matrix into shared memory,
Which, our by hand analysis of the pseudoscope algorithm implies is data-race free, therefore
we expected all data-races checks to return that no data-races occurred.
That was the case for all derived ISCC representations of the algorithm that we were able to generate.





  \newpage

\section{Conclusion}

This case study was overall a success! 
In terms of it being both a learning opportunity for those involved 
and a first step test of a basic idea for creating a data-race/dependency checker for 
SIMT/GPU code, using some common intermediaries.

It is important to note however that the results are more so a proof of concept and would require far more
testing, and proper programmatic implementation before we are comfortable saying that this
is a generally viable methodology for data-race/dependency checking/verification for SIMT/GPU kernel-code.



  \clearpage    
  \pagenumbering{Alph}


\section*{Acknowledgments}
\addcontentsline{toc}{section}{Acknowledgments}

Thank you to Prof Ganesh Gopalakrishnan PhD, for introducing ``us'' to the idea of GPU verification via 
parameterized integer sets and presburger notation, as well as first introducing ``us'' to verification
and GPU/SIMT parallel programing.

Thank you to Tanmay Tirpankar\footnote{\url{https://github.com/tanmaytirpankar}} for providing periodic sanity checks on ``our'' thoughts and work.
They were much needed.


  \addcontentsline{toc}{section}{Refrence}
  \markright{Refrence}
  \printbibliography

  \newpage

\markright{Appendix}
\section*{Appendix}
\addcontentsline{toc}{section}{Appendix}

\subsection*{Reference Code/Listings}
\addcontentsline{toc}{subsection}{Refrence Code/Listings}
This section of the appendix holds the code snippets that could not fit inline with the rest of the text.

\begin{lstlisting}[language=python,%
                  numbers=left,%
                  label={lst:iscc-pollyp-full},%
                  caption={%
                  The full ``code'' for the ISCC representaiton and dependency check for a simple polyhedral product.%
                }]
### Example code (analysed by pet)
# void polynomial_product(int n, int *A, int *B, int *C) {
#     for(int k = 0; k < 2*n-1; k++)
# S:      C[k] = 0;
#     for(int i = 0; i < n; i++)
#         for(int j = 0; j < n; j++)
# T:          C[i+j] += A[i] * B[j];
# }

Domain := [n] -> {
    S[k] : k <= -2 + 2n and k >= 0;
    T[i, j] : i >= 0 and i <= -1 + n and j <= -1 + n and j >= 0;
};

Read := [n] -> {
    T[i, j] -> C[i + j];
    T[i, j] -> B[j];
    T[i, j] -> A[i];
} * Domain;

Write := [n] -> {
    S[k] -> C[k];
    T[i, j] -> C[i + j];
} * Domain;

Schedule := [n] -> {
    T[i, j] -> [1, i, j];
    S[k] -> [0, k, 0];
};

print "Schedule";
print Schedule;

### Happens-Before relation without syntactic sugar
# Lexico := { [i0,i1,i2] -> [o0,o1,o2] : i0 < o0 or i0 = o0 and i1 < o1 or i0 = o0 and i1 = o1 and i2 < o2 };
# Before := Schedule . Lexico . (Schedule^-1)
Before := Schedule << Schedule;

print "Before";
print Before;

RaW := (Write . (Read^-1)) * Before;
WaW := (Write . (Write^-1)) * Before;
WaR := (Read . (Write^-1)) * Before;

print "RaW deps";
print RaW;

print "WaW deps";
print WaW;

print "WaR deps";
print WaR;

IslSchedule := schedule Domain respecting (RaW+WaW+WaR) minimizing RaW;
IslSchedule := IslSchedule + {}; # flatten the schedule tree

print "IslSchedule";
print IslSchedule;

IslBefore := IslSchedule << IslSchedule;

print "IslBefore";
print IslBefore;

print "Does IslSchedule respects RaW deps?";
print RaW <= IslBefore;

print "Does IslSchedule respects WaW deps?";
print WaW <= IslBefore;

print "Does IslSchedule respects WaR deps?";
print WaR <= IslBefore;

print "Codegen";
codegen (IslSchedule * Domain);
\end{lstlisting}
\newpage
\begin{lstlisting}[language=python,%
                  numbers=left,%
                 label={lst:iscc-ge-spmm-full},%
                 caption={%
                  The full ``code'' for the ISCC representaiton of the GeSpMM Algorithm 2.%
                 }]
# ISCC representing Algorithm 2 from:
#         "Ge-SpMM: ..." by Huang et al. - 2020 
#           (https://arxiv.org/pdf/2007.03179.pdf)

Domain := [M,N,K,A_S] -> {
    I[i, j] : 0 <= i < M and 0 <= j < N;
    F[i, j, rs, re, ptr] : 0 <= i < M 
                       and 0 <= j < N 
                       and 0 <= rs <= ptr 
                       and ptr <= re <= A_S;
    T[i, j, rs, re, ptr, kk] : 0 <= i < M 
                           and 0 <= j < N 
                           and 0 <= rs <= ptr 
                           and ptr <= re <= A_S 
                           and 0 <= kk < 32;
};



Read := [M,N,K,A_S] -> {
    I[i, j] -> A_rowPtr[i];
    I[i, j] -> A_rowPtr[i+1];
    F[i, j, rs, re, ptr] -> A_colInd[ptr + (j % 32)] : (ptr + (j % 32)) < re;  
    F[i, j, rs, re, ptr] -> A_val[ptr + (j % 32)] :  (ptr + (j % 32)) < re;     
    T[i, j, rs, re, ptr, kk] -> sm_k[(j - (j % 32)) + kk] :  (ptr+kk) < re;
    T[i, j, rs, re, ptr, kk] -> sm_val[(j - (j % 32)) + kk] :  (ptr+kk) < re;
    T[i, j, rs, re, ptr, kk] -> B[k,j] : 0 <= k < K 
                                     and (ptr+kk) < re;
    T[i, j, rs, re, ptr, kk] -> C[i,j] : (ptr+kk) < re;
  } * Domain;

Write := [M,N,K,A_S] -> {
    F[i, j, rs, re, ptr] -> sm_k[j] : (ptr + (j % 32)) < re;
    F[i, j, rs, re, ptr] -> sm_val[j] : (ptr + (j % 32)) < re;
    T[i, j, rs, re, ptr, kk] -> C[i,j] : (ptr+kk) < re;
} * Domain;

Schedule := [M,N,K,A_S] -> {
    I[i, j] -> [0, i, j];
    F[i, j, rs, re, ptr] -> [1, i, j, rs, re, ptr]; 
    T[i, j, rs, re, ptr, kk] -> [2, i, j, rs, re, ptr, kk];
};

print "Schedule";
print Schedule;

Before := Schedule << Schedule;

print "Before";
print Before;

RaW := (Write . (Read^-1)) * Before;
WaW := (Write . (Write^-1)) * Before;
WaR := (Read . (Write^-1)) * Before;

print "RaW deps";
print RaW;

print "WaW deps";
print WaW;

print "WaR deps";
print WaR;

IslSchedule := schedule Domain respecting (RaW+WaW+WaR) minimizing RaW;
IslSchedule := IslSchedule + {}; # flatten the schedule tree

print "IslSchedule";
print IslSchedule;

IslBefore := IslSchedule << IslSchedule;

print "IslBefore";
print IslBefore;

print "Does IslSchedule respects RaW deps?";
print RaW <= IslBefore;

print "Does IslSchedule respects WaW deps?";
print WaW <= IslBefore;

print "Does IslSchedule respects WaR deps?";
print WaR <= IslBefore;

print "Codegen";
codegen (IslSchedule * Domain);
\end{lstlisting}

\lstlistoflistings
\addcontentsline{toc}{subsection}{List of Listings}

\addcontentsline{toc}{subsection}{List of Figures}
\listoffigures

\addcontentsline{toc}{subsection}{List of Tables}
\listoftables



\end{document}